\documentclass[aps,prd,
nofootinbib,
preprintnumbers,
twocolumn,
showpacs, 
floatfix
]{revtex4-1}
\usepackage{amsmath}
\usepackage{amssymb}
\usepackage{epsfig}
\usepackage{graphicx}
\usepackage{multirow}
\usepackage{color}
\usepackage[normal]{subfigure}
\usepackage{rotating}
\usepackage{hyperref}

\newcommand{\QQ}{{\mathcal{Q}}}

\begin{document}

\thispagestyle{empty}

\preprint{BI-TP 2012/06, ULB-TH/12-06}

\title{Hairy Black Holes in Massive Gravity: Thermodynamics and Phase Structure}

\author{Fabio Capela}
\affiliation{Service de Physique Th\'{e}orique, 
Universit\'{e} Libre de Bruxelles (ULB),\\CP225 Boulevard du 
Triomphe, B-1050 Bruxelles, Belgium}
\author{and Germano Nardini}
\affiliation{Fakult\"at f\"ur Physik, Universit\"at Bielefeld,\\
Universit\"atsstra\ss e 25,
    D-33615 Bielefeld, Germany\\}

  \begin{abstract}The thermodynamic properties of a static and
    spherically symmetric hairy black hole solution
      arising in massive gravity with spontaneous Lorentz breaking
    are investigated. The analysis is carried out by enclosing the
    black hole in a spherical cavity whose surface is maintained at a
    fixed temperature $T$.  It turns out that the ensemble is
    well-defined only if the ``hair" parameter $Q$ characterizing the
    solution is conserved. Under this condition we compute some
    relevant thermodynamic quantities, such as the thermal energy and
    entropy, and we study the stability and phase structure of the
    ensemble.  In particular, for negative values of the hair
    parameter, the phase structure is isomorphic to the one of
    Reissner-Nordstrom black holes in the canonical
    ensemble. Moreover, the phase-diagram in the plan ($Q,T$) has a
    line of first-order phase transition that at a critical value of
    $Q$ terminates in a second-order phase transition. Below this line
    the dominant phase consists of small, cold black holes that are
    long-lived and may thus contribute much more to the energy density
    of the Universe than what is observationally allowed for radiating
    black holes.
\end{abstract}

\maketitle

\section{Introduction}

The discovery of the accelerated expansion of the Universe
\cite{Perlmutter:1998np} has revived interest in the theories of
gravity that can explain such observation without invoking dark
energy.  One class of these models is called ``massive gravity''.
Models of massive gravity fulfilling Lorentz
  invariance have been recently constructed \cite{deRham:2010kj} and
  proved to be free from ghosts and instabilities at full
  non-perturbative level \cite{Hassan:2011hr,deRham:2011rn}. However,
there are very few known solutions for such models.
Instead, models of massive gravity with Lorentz
  violation are better understood at present. In these scenarios the
  spontaneous breaking of Lorentz symmetry is achieved by scalar
fields that are coupled to gravity in a covariant way through
derivative couplings~\cite{ArkaniHamed:2002sp, ArkaniHamed:2003uy,
  key-1}. As a consequence of this interaction, Lorentz violation is
transmitted to the gravitational sector and the graviton becomes
massive~\cite{ArkaniHamed:2003uy,key-1,Rubakov:2004eb}. In
  a wide region of the parameter space these Lorentz breaking models
  are, at perturbative level, free from ghosts and tachyonic
  instabilities around Minkowski \cite{Rubakov:2004eb} and curved
backgrounds \cite{Blas:2009my}. They can also exhibit
  infrared modifications of the gravitational behavior
  \cite{Blas:2009my}. Moreover, they
  are expected to reach the strong coupling regime at energies that
  are much higher than in Lorentz-invariant models of massive gravity
  \cite{ArkaniHamed:2003uy,key-1,Rubakov:2004eb,Blas:2009my}, even though it has not yet been rigorously proved. 
Interestingly, these Lorentz breaking theories are
formulated in a non-perturbative way, making the study of black holes
possible
\cite{Dubovsky:2007zi,Pilo:2008,Bebronne:2009mz,arXiv:1102.0479}. On top of that,
black hole solutions are far richer than in General Relativity (GR)
due to the presence of ``hair" parameters \cite{Pilo:2008,Bebronne:2009mz}.

In GR the existence of black hole solutions raises questions of
consistency with the general laws of thermodynamics. For instance, it
has been argued~\cite{key-0.1} that black holes have
temperature and entropy. The same conclusion was apparently reached  by means of path integral
methods~\cite{Gibbons:1976ue}. However, the proof turned out not to be fully
consistent because of a technical subtlety: the path integral approach
requires black holes in thermal equilibrium with its surroundings and
this situation is never fulfilled by Schwarzschild
solutions. Instead, anti-de-Sitter
(AdS) black holes achieve thermal equilibrium and the path integral
procedure proposed in Ref.~\cite{Gibbons:1976ue} can be consistently
applied~\cite{Hawking:1982dh}. Moreover, it turns out that black
holes are just one of the phases permitted in AdS space and first
order transitions between a black hole and globally-AdS spacetime may
occur\,\footnote{In the context of the
  AdS/CFT correspondence \cite{Maldacena:1997re}, this transition
  turns out to be dual to the confinement-deconfinement phase
  transition in large-$N$ gauge theories \cite{Witten:1998zw}.}. On
the other hand, it might be expected that a similar result also arises
for Schwarzschild black holes if one mimics the AdS cosmological
constant by an infrared cutoff.  In fact, as it was proven by
York~\cite{York:1986it}, after having enclosed the Schwarzschild black
hole inside a spatial spherical surface at a fixed temperature, the
configuration reaches thermal equilibrium and its phase
structure turns similar to the AdS one.

It is interesting to understand whether one can apply the York's
approach to investigate thermodynamics of hairy black hole solutions in 
Lorentz breaking massive
gravity. This issue is the subject of the present paper and, as we
will see, the conclusion depends on whether the black hole ``hair''
parameter $Q$ is conserved in the cavity. In particular, when $Q$ is
assumed constant~\footnote{In practice, we expect that the results
  obtained under this hypothesis can be extended to situations where
  $Q$ varies very slowly.}, we obtain a rich
thermodynamical phase structure that includes, for instance, first and
second order phase transitions and phases consisting of stable
and small, cold black holes (plausible dark matter candidates).

The paper is organized as follows. In Sect.~\ref{sec:1} we briefly
review the model of massive gravity and we sketch the eventual
difficulties that may arise when applying the York's procedure. In
Sect.~\ref{sec:2} we review the static, spherically symmetric and
asymptotically-flat black hole solution of massive gravity and we
check whether it can be embedded in a spacetime that has a periodic
Euclidean time coordinate and is bounded by a spatial sphere.  This
compatibility requirement is needed in order to apply the York's
procedure, which we carry out in Sect.~\ref{sec:3} assuming $Q$ to be
conserved in the ensemble. In particular, we first regularize the
black hole action by the ``subtraction background"
method~\cite{Gibbons:1976ue}, then we enclose the black hole solution
in a sphere at a fixed temperature and, finally, we obtain the
on-shell action of the ensemble. As we prove in Sect.~\ref{sec:4},
this procedure provides a well-defined ensemble that we use to
investigate black hole thermodynamics. It turns out that, depending on
the value of $Q$ inside the cavity, black holes can evolve in
qualitatively different ways. In any case, their evolution always
respects the Bekenstein-Hawking formula.  In Sect.~\ref{sec:5} we
relax the assumption concerning the conservation of $Q$ and we see that
in such a case the York's procedure is inconsistent. Finally,
Sect.~\ref{sec:6} is dedicated to summarize the main results of the
paper and Appendix contains some analytic expressions valid for a
specific choice of massive gravity parameters.

\section{Massive gravity and the saddle-point approximation} \label{sec:1}
Massive gravity is conventionally described by the action
\cite{key-1,Rubakov:2004eb, Blas:2009my, Dubovsky:2007zi,Pilo:2008,Bebronne:2009mz, arXiv:1102.0479}
	\begin{eqnarray}
		I&=& \int_{\mathcal{M}} d^4 x \sqrt{g} \left[- \frac{1}{16 \pi} R+ \Lambda^4 \mathcal{F}(X,W^{ij}) \right]\nonumber\\ &-&  \int_{\partial \mathcal{M}} d^3 x \sqrt{\gamma} \; \frac{1}{8\pi}K  \label{eq:1}~,
	\end{eqnarray}
with
\begin{eqnarray}
X&=&\Lambda^{-4}g^{\mu\nu} \partial_\mu
\phi^0 \partial_{\nu}\phi^0~,\nonumber\\ V^i&=&\Lambda^{-4}\partial ^{\mu}
\phi^i \partial_{\mu} \phi^0~,\nonumber\\
W^{ij}&=&\Lambda^{-4}\partial^{\mu}\phi^i \partial_{\mu} \phi^{j}-
\frac{V^i V^j}{ X}~,\nonumber
\end{eqnarray}
where Latin (Greek) indices run on space (spacetime)
components. The first integral in Eq.~\eqref{eq:1} is evaluated on the
manifold $\mathcal{M}$ with metric $g$ and contains two contributions:
the usual Einstein-Hilbert term and a function $\mathcal{F}$ of four scalar
fields $\phi^{\mu}$ that are minimally coupled to gravity by covariant
derivatives. The second integral is instead the Gibbons-Hawking-York
boundary term \cite{Gibbons:1976ue,York:1972sj}, where $\gamma_{ab}$
is the metric induced on the boundary $\partial \mathcal{M}$ and $K$
is the trace of the extrinsic curvature
$K_{ij}=\frac{1}{2}\gamma^k_{i}\nabla_k n_j$ of $\partial \mathcal{M}$
with unit normal $n^i$. Such a boundary term is required to have a
well-defined variational principle in the presence of the border
$\partial\mathcal{M}$.

The action \eqref{eq:1} describes a low-energy effective theory valid
below the ultraviolet cutoff $\Lambda$, which
  perturbative analyses estimate to be ${\mathcal O}(\sqrt{m
    M_{\text{Pl}}})$
  \cite{ArkaniHamed:2003uy,key-1,Rubakov:2004eb,Blas:2009my}, where
$M_{\text{Pl}}=\sqrt{1/8\pi}$ and $m$ are the Planck and the graviton
masses, respectively.  Its vacuum flat-spacetime solution has the form
	\begin{equation}
		g_{\mu\nu}=\eta_{\mu\nu}~,~\qquad \phi^0_{\text{flat}}=\Lambda^2 t~,~\qquad \phi^i_{\text{flat}}=\Lambda^2 x^i~,~ \label{eq:2}
	\end{equation}	
which induces a spontaneous breaking of Lorentz symmetry. 	
The background \eqref{eq:2} preserves rotational symmetry when the function
$\mathcal{F}$ is invariant under rotations in the internal space of the fields
$\phi^i$. Moreover, the action \eqref{eq:1} is invariant under the symmetry
	\begin{equation}
		\phi^i\rightarrow \phi^i +\Theta^i\left(\phi^0 \right)~,	\label{eq:symm}
	\end{equation}
where $\Theta^i$ are arbitrary functions of $\phi^0$. This symmetry ensures
that perturbations around the vacuum contain only two propagating degrees of
freedom \cite{key-1}, corresponding to the two polarizations of a massive
graviton.

To study the thermodynamics of this model, we use the Euclidean path integral 
	\begin{equation}
		\mathcal{Z}= \int \mathcal{D} g \mathcal{D} \phi \; \exp\left( - I\left[g,\phi\right]\right)~,\label{eq:Z}
	\end{equation}
which is evaluated by integrating over all metrics and scalar fields
satisfying particular boundary conditions.  In the semi-classical limit
$\mathcal{Z}$ is dominated by the stationary points of the action. This can be
checked by expanding the path integral around a classical solution.  Indeed,
if the expansion provides a leading term that is finite, a linear term that
vanishes on-shell, and a quadratic term that is positive definite, then the
function $\mathcal{Z}$ can be expressed as
\begin{equation}
  \mathcal{Z}\approx\text{e}^{-I[g_{\text{cl}},\phi_{\text{cl}}]} \int \mathcal{D} \delta g \mathcal{D} \delta \phi \;\;\text{e}^{- \delta^2 I\left[g_{\text{cl}},\phi_{\text{cl}};\delta g,\delta \phi\right]} \label{eq:2.5}
\end{equation}
and can be interpreted as the partition function of the model.  However,
such a derivation is not straightforward for the action \eqref{eq:1} since the
three properties listed above might not be fulfilled. In fact:
\begin{enumerate}
\item The on-shell leading term of the action diverges. This is a
  familiar problem in general relativity that is addressed using the
  ``background subtraction'' regularization technique
  \cite{Gibbons:1976ue}. Following this subtraction scheme, we take
  the vacuum solution \eqref{eq:2} as background $(g_0,\phi_0)$ and we consider $I_E$ as regularized action, defined
  as
	\begin{equation}
		I_E(g,\phi)\equiv I(g,\phi)-I(g_0,\phi_0)~.
\label{Ireg}
\end{equation}
In this way $I_E$ is finite for the class of fields $(g,\phi)$ that
asymptotically approach the background $(g_0,\phi_0)$;

\item The linear term may eventually not vanish for all perturbations around
  the classical solution.  The non-vanishing behavior of such a term comes
  from the boundary contributions:
	\begin{equation}
		\delta I \left|\right._{\text{cl}} = \int _{\partial
                  \mathcal{M}} d^3x\sqrt{\gamma}\;
                \left[\pi^{ab}\delta\gamma_{ab}+\pi^{\phi}_{\mu}\delta\phi^\mu\right]~.
\end{equation}
For the action \eqref{eq:1} $\pi^{ab}$ and $\pi^{\phi}_{\mu}$ are
given by
	\begin{eqnarray}
		&&\pi^{ab}= -\frac{1}{16\pi}\left(K^{ab}-\gamma^{ab}K\right)~,\label{eq:3_1}\\
		&&\pi^{\phi}_{\mu}=2\Lambda^4n_{\alpha} \left[\left(\frac{\partial \mathcal{F}}{\partial X}
		+\frac{\partial \mathcal{F}}{\partial W^{ij}} \frac{V^i V^j}{X^2}\right)\delta_{\mu}^0 \partial ^\alpha \phi^0 \right. \nonumber\\	
		&&\left.+\frac{\partial \mathcal{F}}{\partial W^{ij}}\delta_{\mu}^i \partial ^{\alpha} \phi^j- \frac{\partial \mathcal{F}}{\partial W^{ij}}\frac{V^j}{X}\partial^{\alpha} \left(\phi^0 \delta_{\mu}^i+\phi^i\delta_{\mu}^0\right)\right]~,
			\label{eq:3}
\end{eqnarray}
where $n_{\alpha}$ is the outward pointing unit normal to $\partial
\mathcal{M}$. Therefore, some boundary conditions on the fields $g_{\mu\nu}$
and $\phi^{\mu}$ should be imposed in order to have a well-posed variational
problem, i.e.~$\left.\delta I \right |_{\text{cl}}=0$;

\item The Gaussian integral in Eq.~\eqref{eq:2.5} corresponds to the one-loop contribution. 
Such contribution may diverge since the integration involves the determinant of second-order
elliptic operators that cannot be regularized when negative eigenvalues are present.
Indeed, if there exist
negative eigenvalues, the density of states grows so rapidly that the ensemble
turns out to be ill-defined and
$\mathcal{Z}$ cannot be used to determine thermodynamical quantities. For this
reason, in order to be able to study the thermodynamical properties of black
holes, we need first to stabilize the ensemble. This can be performed as in
Ref.~\cite{York:1986it}: we place the black hole inside a surface
maintained at a fixed temperature.

\end{enumerate}

In the next  sections we explicitly show how to implement these
three procedures for black hole solutions in massive gravity. 

\section{The black hole solutions}\label{sec:2}

By the coordinate transformations $r'\rightarrow r=r(r')$ and
$t'\rightarrow t=t+\tau(r')$, the generic ansatz for the static
spherically symmetric solution in Euclidean spacetime can be written
as \cite{Pilo:2008,Bebronne:2009mz}
\begin{eqnarray}
		ds^2&=& \alpha(r)dt^2+\rho(r)dr^2+r^2\left(d\theta^2+\sin^2 \theta d\varphi^2\right)~, \nonumber\\
		\phi^0&=& \Lambda^2\left[-it+h(r)\right]~,\label{eq:ans1} \\
		\phi^i&=& \phi(r)\frac{\Lambda^2 x^i}{r}~. \nonumber 
\end{eqnarray}
For black hole solutions in massive gravity the explicit expression of this
ansatz can be obtained by imposing Eqs.~\eqref{eq:ans1} to fulfill  the
equations of motions of the action~\eqref{eq:1}. The black hole solution has then to be an
extremum of the variation of the action with respect to the fields
$g_{\mu\nu}$ and $\phi^{\mu}$:
	\begin{eqnarray}
		\delta I&=& \int_{\mathcal{M}} d^4x
                \sqrt{g}\left[\textbf{E}^{\mu\nu}_{(1)}\delta
                  g_{\mu\nu}+\textbf{E}_{\mu}^{(2)}\delta\phi^{\mu}
                \right]\nonumber\\&+&\int _{\partial \mathcal{M}}
                d^3x\sqrt{\gamma}\;
                \left[\pi^{ab}\delta\gamma_{ab}+\pi^{\phi}_{\mu}\delta\phi^\mu\right]=0~,
\label{eq:dI}
\end{eqnarray}
where 
	\begin{eqnarray}
		&&\textbf{E}^{\mu\nu}_{(1)}=-\frac{1}{16\pi}\left(R^{\mu\nu}-\frac{1}{2}Rg^{\mu\nu}-8\pi T^{\mu	\nu}_{\phi}\right)~, \nonumber\\ &&T_{\phi}^{\mu\nu}= \frac{2 \Lambda^4}{\sqrt{g}}\frac{\delta\left(\sqrt{g}\mathcal{F}(X,W^{ij})\right)}{\delta g_{\mu\nu}}~, \\
		&&\textbf{E}_{\mu}^{(2)} =-2\Lambda^4\nabla_{\alpha}\left[\left(\frac{\partial \mathcal{F}}{\partial X}
		+\frac{\partial \mathcal{F}}{\partial W^{ij}} \frac{V^i V^j}{X^2}\right)\delta_{\mu}^0 \partial ^\alpha \phi^0 \right. \nonumber\\	
		&&\left.+\frac{\partial \mathcal{F}}{\partial W^{ij}}\delta_{\mu}^i \partial ^{\alpha} \phi^j- \frac{\partial \mathcal{F}}{\partial W^{ij}}\frac{V^j}{X}\partial^{\alpha} \left(\phi^0 \delta_{\mu}^i+\phi^i\delta_{\mu}^0\right)\right]~.\nonumber
\end{eqnarray}
For asymptotically-flat black holes the solution of Einstein and Goldstone's
equations, respectively  $\textbf{E}^{\mu\nu}_{(1)}=0$ and $\textbf{E}^{(2)}_{\mu}=0$, has a
known analytical expression if the function $\mathcal{F}$ takes the
form~\cite{Bebronne:2009mz}
	\begin{equation}
		\mathcal{F}=\frac{12}{\lambda X} +6 \left(\frac{2}{\lambda}+1\right)w_1-w_1^3+3w_1w_2-2w_3+12~,
		\label{eq:ans2}
\end{equation}
where $\lambda$ is a positive constant and $w_n=\text{Tr}(W^n)$. In
such a case, for $\lambda \neq 1$ , the ansatz \eqref{eq:ans1}
provides the black hole solution
\begin{eqnarray}
  \alpha(r) &=& 1-\frac{2M}{r}-\frac{Q}{r^{\lambda}}~,\nonumber\\
  \rho(r) &=& \frac{1}{\alpha(r)}~,  \label{eq:sol} \\ 
  h(r) &=& \pm \int \frac{dr}{\alpha}\left[1-\alpha\left(\frac{Q}{12m^2}\frac{\lambda(\lambda-1)}{r^{\lambda+2}}+1\right)^{-1}\right]^{1/2}~,\nonumber\\
  \phi(r) &=& r~,  \nonumber
\end{eqnarray}
which depends on the two arbitrary integration constants $M$ and
$Q$. In the following we restrict our analysis to the class of
solutions \eqref{eq:sol} with $\lambda> 1$, so that the gravitational
potential is asymptotically Newtonian and the parameter $M$ coincides
with the Arnowitt-Deser-Misner (ADM) mass \footnote{These features have been discussed in Ref.~\cite{Bebronne:2009mz}
 for $\lambda<1$. For the particular case $\lambda=1$,
  yielding a different solution from \eqref{eq:sol}, see Appendix F of
  Ref.~\cite{Pilo:2008}. Such a solution produces a divergent ADM mass .}.
  ~Moreover, we forbid naked singularities, i.e. $\alpha(r)$
must have real roots and the largest of them determines the radius of
the event horizon. Depending on the signs and relative values of the
parameter $M$ and $Q$, the following cases arise:

\begin{itemize}

\item{ $Q\ge 0$ and $M>0$ :} The existence of the horizon is
  guaranteed: at all distances the gravitational potential is attractive and
  stronger than (or, for $Q= 0$, equal to) the usual Schwarzschild black hole
  potential. Therefore $r_+$ is never smaller than in the standard case.

\item{$Q\ge0$ and $M<0$ :} The Newton's potential is repulsive
  at large distances and attractive near the horizon. This possibility is
  interesting because, even for $Q=0$, it does not have a
  corresponding case in GR\,\footnote{The choice $M<0$ is problematic in
    GR where: {\it i)} it leads to naked singularities; {\it ii)} it violates
    the null energy condition $T_{\mu\nu}k^{\mu}k^{\nu}\geq0$ (being $k^\mu$ a
    future-pointing null vector field), which holds for the matter stress
    tensor \cite{key-8} and implies the positivity of the
    ADM mass. Neither of the arguments exist in massive gravity: {\it i)} at short distance the
    repulsion changes to attraction, which creates the event horizon; {\it ii)}
    the stress-energy tensor  of the scalar fields $T_{\mu\nu}^\phi$ does not satisfy 
    the null energy condition, allowing for negative mass states to be constructed, e.g., as in the ghost condensate model~\cite{key-11}.}.
 Nonetheless, we do not analyze such negative mass configurations
  since they are incompatible with black hole thermodynamical laws
  \cite{arXiv:1102.0479}.  As stated in
  Ref.~\cite{Feldstein:2009qy}, it seems likely that the cause of
  these incompatibilities is not Lorentz violation but the
  existence of negative energy states.

\item{ $Q<0$ and $M>0$ :} The horizon only exists when the condition
\begin{equation}
		2M\geq \lambda |Q|^{1/\lambda} \left(\frac{1}{\lambda-1}\right)^{\frac{\lambda-1}{\lambda}}\label{eq:cond}
\end{equation}
is fulfilled.
In this case the Newton's potential is always attractive until reaching the
horizon but the attraction is weaker (which makes the event horizon radius $r_+$ smaller) than
in the Schwarzschild case. 

\end{itemize}

\subsection{The boundary conditions}
As it has already been mentioned in Sect.~\ref{sec:1}, the procedure to study
equilibrium thermodynamics requires to enclose the asymptotically-flat black
hole within a finite volume  surface and then send the surface to
infinity~\cite{York:1986it} (see Sect.~\ref{sec:3} for details). In the
following, we consider a spherical cavity of radius $r_\mathcal{B}$ as
the boundary and, to analyze the system at finite temperature, we impose
periodicity on the Euclidean time. Due to the presence of this border, the
black hole solution \eqref{eq:sol} must fulfill some boundary conditions in
order to satisfy $\delta I|_{\text{cl}}=0$.
The momentum $\pi^{\phi}_0$ conjugate to the scalar field $\phi^0$
on $\partial \mathcal M$ vanishes when evaluated at the solution
\eqref{eq:sol}. Then, there is no need to fix the scalar field
$\phi^0$ at the boundary. Analogously, as a remnant of the internal
spherical symmetry of the scalar fields $\phi^i$, i.e. $(\phi^i)^2|_{\partial\mathcal{M}}=r_{\mathcal{B}}$, 
no boundary conditions emerge on the scalar sector.
Instead, the momenta
$\pi^{ab}$ conjugated to the induced metric $\gamma_{ab}$ are not null
when evaluated at the background solution \eqref{eq:sol}. For this
reason we conclude that we only need to impose
$\delta\gamma_{ab}=0$ to have
vanishing boundary terms.
	\begin{figure} 
	 \includegraphics[scale=0.8]{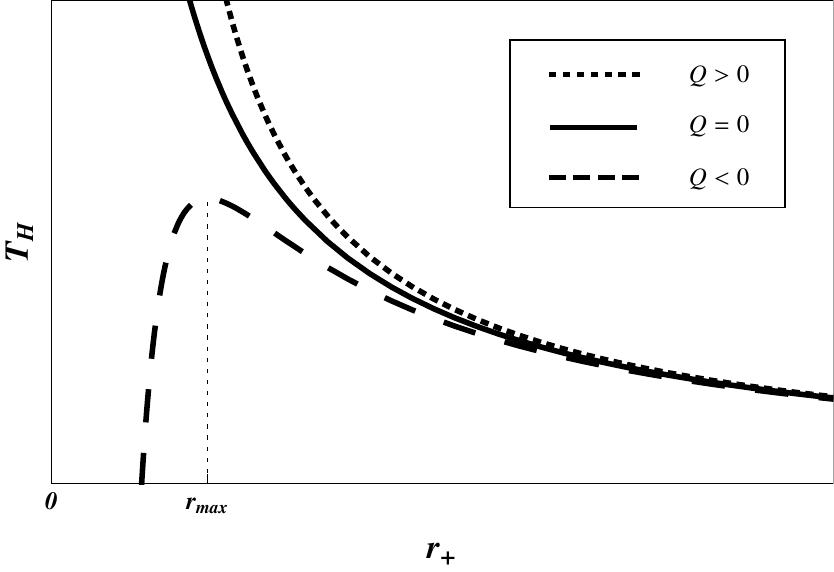}
	      \caption{The Hawking temperature $T_H$ as a function of the
                event horizon radius $r_+$ for positive and negative scalar
                charge $Q$ and for the conventional Schwarzschild case
                $Q=0$. For $Q<0$ the Hawking temperature reaches a maximum at
                $r_+=r_{\text{max}}\equiv\left[(\lambda^2-1)|Q|\right]^{1/\lambda}$.}
\label{fig:1}
\end{figure}
Further conditions arise due to the compactification of
the Euclidean time. Regularity of the metric at the horizon fixes the
periodicity as $t\sim t+\beta_H$~, where $\beta_H$ is
related to the Hawking temperature $T_H$ by the equality
	\begin{equation}
          T_H^{-1}= \beta_H  = \left.\frac{4\pi}{\partial_r \alpha} \right|_{r_+}=\frac{4\pi r_+} {1+(\lambda-1)\frac{Q}{r_+^\lambda}}~.\label{eq:3.6}
\end{equation}
Interestingly, the Hawking temperature behaves differently for positive and
negative scalar charges (see Fig.~\ref{fig:1}). For $Q>0$ the temperature is
larger than for $Q=0$ and decreases with the increasing of the event horizon radius $r_+$,
as in the case of the conventional Schwarzschild black hole. For $Q<0$ the temperature
is smaller than in the Schwarzschild case and its non-monotonic dependence on
$r_+$ reaches a maximum at
$r_+=r_{\text{max}}\equiv\left[(\lambda^2-1)|Q|\right]^{1/\lambda}$.  The
periodicity of $t$ also imposes a periodicity on the scalar field $
\phi^0(t,r)\sim\phi^0(t+\beta_H,r)$ which is linearly dependent on time. This
does not provide any further constraint on the solution \eqref{eq:sol} since
the periodicity of $\phi^0$ is allowed due to the presence of the global
shift symmetry $\phi^0\rightarrow \phi^0 +C$ in the action \eqref{eq:1}.

Notice that the Hawking temperature measured by an observer at the position
$r$ is given by
\begin{equation}
		T(r)=\frac{\beta^{-1}_H}{\sqrt{\alpha(r)}}= \frac{1}{4\pi r_+} \frac{ \left(1+(\lambda-1) \frac{Q}{r_+^{\lambda}}\right)}{\sqrt{1-\frac{r_+}{r}+\frac{Q}{r_+^{\lambda-1}r}-\frac{Q}{r^{\lambda}}}}~.\label{eq:T}
\end{equation}
In case of thermal equilibrium, setting the temperature measured at the
surface $r=r_{\mathcal B}$ univocally determines the temperature of the
configuration contained in the cavity. Once one has set $r_\mathcal{B}$
and $T(r_\mathcal{B})$ the boundary $\partial \mathcal{M}$ is unambiguously fixed.

\section{The on-shell action} 
\label{sec:3} 
In this section we briefly review the background subtraction procedure needed
to obtain a finite on-shell action. Subsequently, we compute the Euclidean
action of the black hole solutions \eqref{eq:sol} in terms of the boundary
conditions $(T(r_{\mathcal{B}}),r_\mathcal{B})$ under the assumption that $Q$
is a conserved quantity in the cavity.

\subsection{The regularization procedure} 

A priori, the Euclidean action \eqref{eq:1} may present an integration problem
at $r=r_+$ and another at $r\rightarrow \infty$. The former is solved as usual by
assigning the period $\beta_{H}$ to $t$, as in such a case the metric
\eqref{eq:sol} extends smoothly onto the event horizon. The latter is instead
more cumbersome. To regularize it~\cite{York:1986it}, we perform the integration from $r=r_+$
up to the infrared cutoff $r=r_\mathcal{B}$, we subtract off the action of the
vacuum flat space \eqref{eq:2}\,\footnote{Of course, this subtracted
  background is not $\mathbb R^4$ but $\mathrm S^1\times \mathbb R^3$ since
  the periodicity of time is maintained.}, and finally we send $r_\mathcal{B}$
to infinity. In this way the Euclidean action is regularized and defined as in
Eq.~\eqref{Ireg}.

Notice that in a spherical cavity of finite volume the on-shell action (with
the metric regularized at $r=r_+$) is always finite and, in addition, the
thermodynamic stability is guaranteed~\cite{York:1986it}. Hence, as long $
r_{\mathcal{B}}$ is finite, in principle there is no need to subtract any
background to make the action finite. Instead, this procedure is necessary to have no divergent thermodynamical properties when the limit
$r_{\mathcal{B}}\rightarrow\infty$ is taken. Then, in view of
this limit, we apply the regularization procedure as a first step to analyze
black hole thermodynamics.

\subsection{The regularized action for hairy black holes}
\label{sec:42}
An apparent ambiguity may arise in the procedure to regularize the action
\eqref{eq:1}: since the subtracted background \eqref{eq:2} is regular
everywhere, it does not require any specific periodicity of the time
coordinate. However, black hole and background metrics have to match 
at the boundary surface $r=r_{\mathcal{B}}$. Thus, the time periodicity $\beta$
of the background has to be
	\begin{equation}
		\beta=T(r_{\mathcal{B}})^{-1} 
~,\label{eq:m1}
\end{equation}
where $T(r_{\mathcal{B}})^{-1}=\beta_{H}\sqrt{\alpha(r_{\mathcal{B}})}$ as in
Eq.~\eqref{eq:T}.
\begin{figure*}
 \begin{center}
\hspace{-.6cm} \subfigure{\includegraphics[scale=0.8]{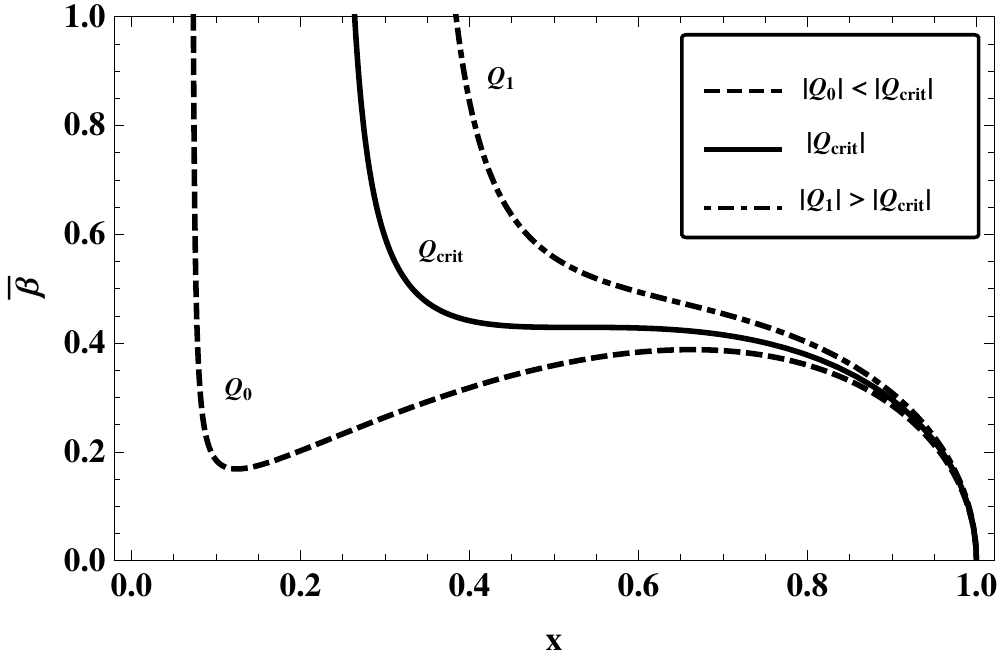}}\hspace{.9cm}
   \subfigure{\includegraphics[scale=0.8]{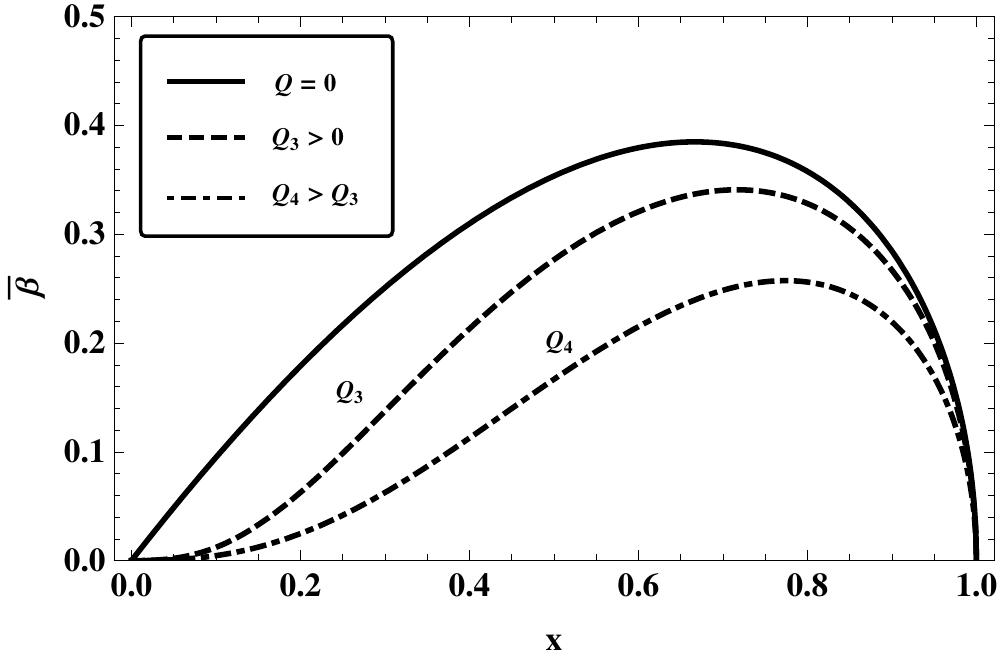}}
    \caption{\underline{Left panel}: $\bar{\beta}$ as a function of the event horizon for
     negative scalar charges. Depending on the value of the scalar charge, the
     number of solutions is one or three. \underline{Right panel}: $\bar{\beta}$
     as a function of the event horizon for positive scalar charge. Depending
     on the value of $\bar{\beta}$, there exist two or no solutions. Negative
     slopes correspond to thermally stable states.}
   \label{fig:2}
   \end{center}
\end{figure*}

Once the action \eqref{eq:1} has been regularized, the evaluation of
$I_E$ for the black hole solution \eqref{eq:sol} with radial
integration $r_+\leq r \leq r_{\mathcal{B}}$ is
straightforward. Indeed, as the metric \eqref{eq:sol} and
$\mathcal{F}$ of Eq.~\eqref{eq:ans2} are stationary, the time integration in $I_E$ just gives
rise to a multiplicative factor $\beta$. Moreover, the integration on
the other variables can be performed in a closed manner. 
The regularized (quasi-local) action of the black hole then results to
be
	\begin{equation}
		I_E=\beta\mathcal{E}_{BY} -\frac{1}{4}\mathcal{A}_{\text{BH}}~,\label{eq:IE}
	\end{equation}
with
	\begin{eqnarray}
	\mathcal{A}&\equiv& 4\pi r_+^2~,\label{eq:E} \\ \mathcal{E}_{BY}&=&
  r_{\mathcal{B}}\left[1-\sqrt{\alpha(r_\mathcal{B})}\right]=\frac{1}{8\pi}\int_{\mathcal{B}_{\bar{t}}}d^2x
\sqrt{\sigma} \left(k-k_0\right)~, \nonumber  
\end{eqnarray}
where $k$ is the trace of the extrinsic curvature of the two-boundary
$\mathcal{B}_{\bar{t}}\equiv\partial \mathcal{M}|_{t=\bar{t}}$\,, $\sigma$ is the induced
metric on $\mathcal{B}_{\bar{t}}$ and $k_0$ refers to the extrinsic curvature of
$\mathcal{B}_{\bar{t}}$ embedded in the vacuum space \eqref{eq:2}. 

The last equality in Eqs.~\eqref{eq:E} shows explicitly that $\mathcal{E}_{BY}$
is the Brown-York quasi-local energy \cite{Brown:1992br}, corresponding to the
Hamiltonian that generates the time translation at the
two-boundary $\mathcal{B}_{\bar{t}}$.  For this reason we can interpret
$\mathcal{E}_{BY}$ in Eq.~\eqref{eq:IE} as the energy of the black
hole\,\footnote{Remind that by construction $\mathcal{E}_{BY}=0$ for the vacuum
  spacetime \eqref{eq:2}.}.  On the other hand, plugging
Eq.~\eqref{eq:sol} in Eq.~\eqref{eq:E} reads
	\begin{equation}
		M=\mathcal{E}_{BY}-\frac{\mathcal{E}_{BY}^2}{2r_{\mathcal{B}}}-\frac{Q}{2r_{\mathcal{B}}^{\lambda-1}}~,
\label{adm}
\end{equation}
showing that the ADM mass $M$ is the total energy of the black hole in
the limit $r_{\mathcal{B}}\rightarrow \infty$.  For this reason, when we send
the cavity surface to infinity, the regularized on-shell action is given by
	\begin{equation}
		I_E=\beta_H M- \frac{1}{4}\mathcal{A}_{\text{BH}}~,\label{eq:onsh}
\end{equation}
which still vanishes when the event horizon goes to zero as it
reproduces the action of the subtracted vacuum spacetime\,\footnote{Of
  course, this conclusion is correct just for $M>0$, as we are
  assuming.}.
 A detailed discussion about the finite energy of the black hole solutions \eqref{eq:sol}
 can be found in Ref. \cite{energy}.

\section{Thermodynamics and phase transitions} \label{sec:4}
Eq.~\eqref{eq:T} evaluated at $r=r_{\mathcal{B}}$ produces a function
of the ADM mass $M$ in terms of the parameters of the ensemble
$\beta$, $r_{\mathcal{B}}$ and $Q$. Depending on the particular values
of these parameters, there can exist zero, one or multiple black hole
solutions that are allowed inside the cavity. When several
configurations are permitted, phase transitions may occur. In the present section we analyze this issue under
the hypothesis that the scalar charge is conserved inside the cavity
(for considerations without this assumption see Sect.~\ref{sec:5}).

\subsection{The phases}
In order to determine the number of black hole solutions allowed inside a
cavity containing a given scalar charge, we work out the temperature at the
boundary, $\beta^{-1}$, as a function of $r_+$ for fixed $Q$.  This can be
done by taking Eq.~\eqref{eq:T}:
	\begin{equation}
		\bar{\beta}(x,\QQ)=x\frac{\sqrt{1-x}\sqrt{1+\frac{\mathcal{Q}}{x^{\lambda-1}}\frac{1-x^{\lambda-1}}{1-x}}}{1+\left(\lambda-1\right) \frac{\mathcal{Q}}{x^{\lambda}}}\label{eq:mast}~,
\end{equation}
where $x\equiv r_+/r_{\mathcal{B}}$, $\mathcal{Q}\equiv
Q/r_{\mathcal{B}}^{\lambda}$ and $\bar{\beta}\equiv\beta/4\pi
r_{\mathcal{B}}$.  Observe that $\bar{\beta}$ positivity is guaranteed for
$Q>0$ and is equivalent to require Eq.~\eqref{eq:cond} for $Q<0$.

The plot of $\bar{\beta}(x,\QQ)$ is presented in Fig.~\ref{fig:2} for specific values of $\QQ$. The curves are obtained for a certain $\lambda$ but the
general features of the solution are independent of this choice.

For the case $Q>0$ (right panel of Fig.~\ref{fig:2}) the function
$\bar{\beta}(x,\QQ)$ exhibits a simple behavior: if
$\bar{\beta}$ is below a certain value $\bar{\beta}_{\text{max}}$, two black
hole configurations with the same $Q$ are allowed; otherwise no black hole (with the assumed scalar
charge) is permitted. Moreover, the value of $\bar{\beta}_{\text{max}}$ gets
larger as $Q$ decreases and it is bounded by the relation
$\bar{\beta}_{\text{max}}\le 2/\sqrt{27}$ where the equality occurs for
Schwarzschild black holes.

For the case $Q<0$ (left panel of Fig.~\ref{fig:2}) $\bar{\beta}$ exhibits
three different qualitative behaviors. For
$|\mathcal{Q}|>|\mathcal{Q}_{\text{crit}}|$ (with
$\mathcal{Q}_{\text{crit}}<0$ quantified later) at each temperature there is
one single solution. In particular, the event horizon of the black hole grows
as the temperature and reaches the boundary only when $\bar{\beta}=0$. For
$|\mathcal{Q}|<|\mathcal{Q}_{\text{crit}}|$ there is an interval of
temperatures outside which $\bar{\beta}$ behaves as in the previous case,
i.e.~only large (nearly extremal) black holes exist at very high (very low)
temperature. Instead, for $\bar{\beta}$ within this interval, three black hole
configurations are allowed. These three solutions approach the same event
horizon for $|\mathcal{Q}|\rightarrow|\mathcal{Q}_{\text{crit}}|^-$ and for
this reason $|\mathcal{Q}_{\text{crit}}|$ can be defined as the value of $\QQ$
at which the extrema of $\bar{\beta}$, located at $\partial \bar{\beta}
/\partial x =0$, coincide. For instance, when $\lambda=2$ it turns out to be
(see Appendix for details)
	\begin{equation}
		|\mathcal{Q}_{\text{crit}}|=\frac{1}{9+4\sqrt{5}}~. \label{eq:ccrit}
\end{equation}

\begin{figure*}
   \begin{center}
\hspace{-.6cm}\subfigure{\includegraphics[scale=0.8]{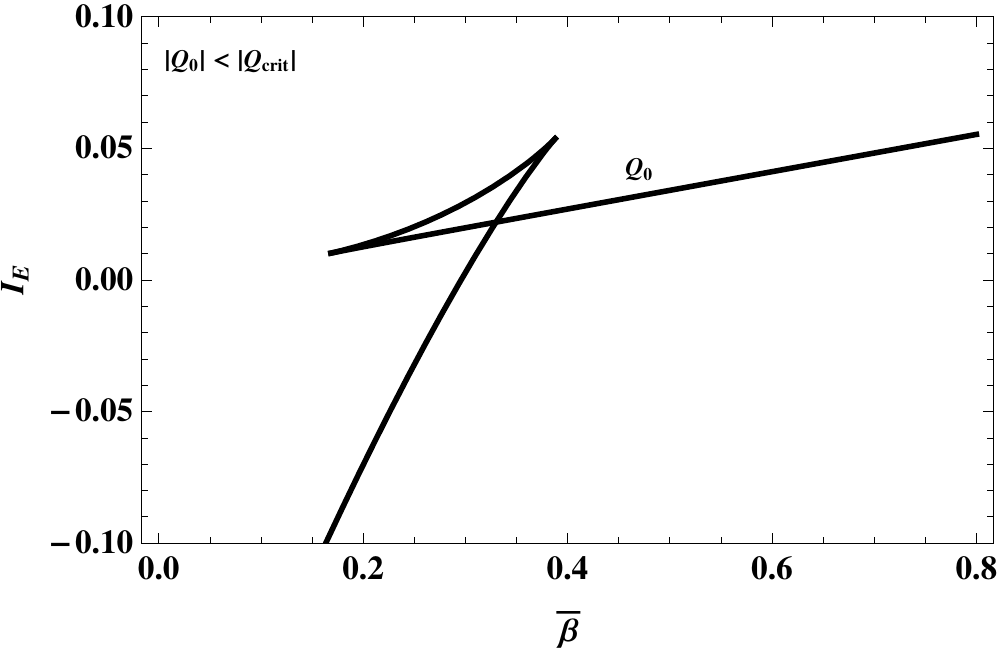}}\hspace{.9cm}
\subfigure{\includegraphics[scale=0.8]{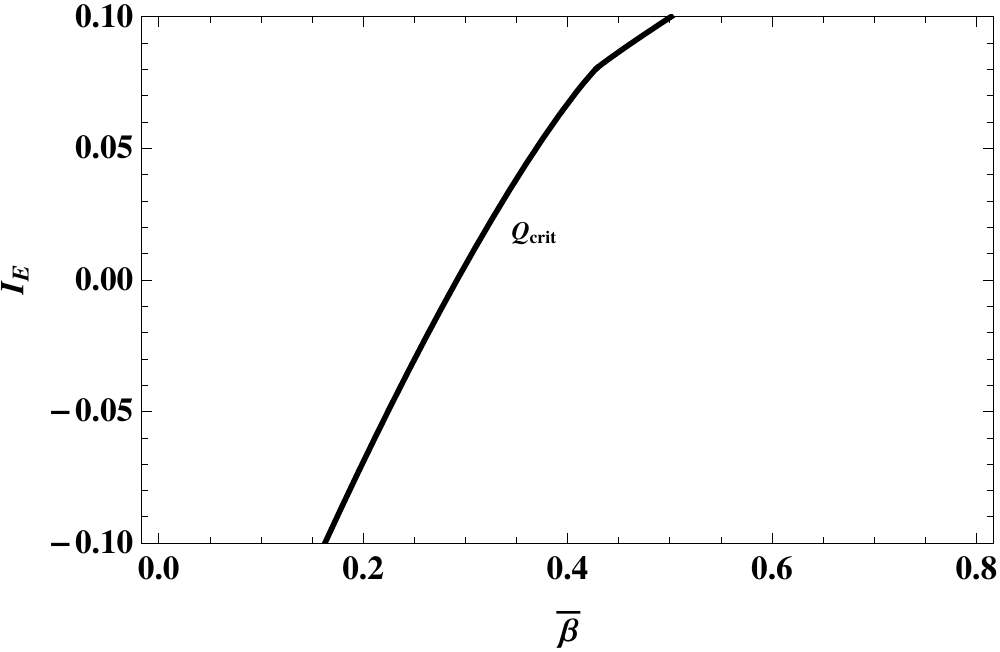}}\\
\hspace{-1.1cm}\subfigure{\includegraphics[scale=0.8]{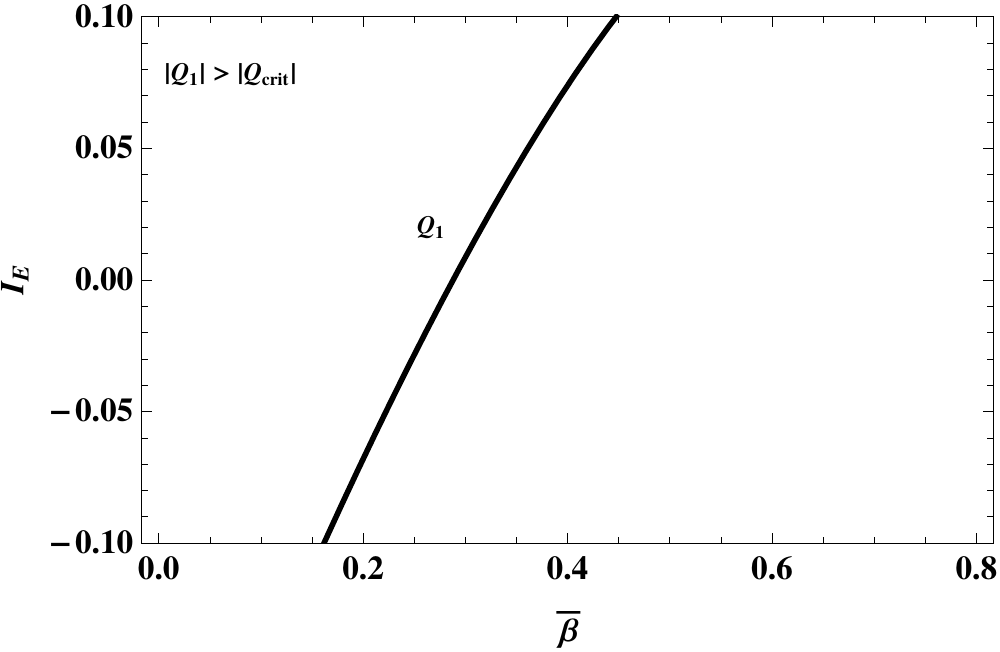}}
   \caption{The Euclidean action as a function of $\bar{\beta}$ for negative
     scalar charge in the three relevant cases. The values of $Q$ are the same
     as in Fig. \ref{fig:2} }
    \label{fig:4}
   \end{center}
\end{figure*}

\subsection{The local stability}\label{sec:local}
The local stability in the saddle point approximation is related to the
convergence of the integral in the on-shell partition function \cite{Brown:1994su}, as we briefly review now.

The on-shell partition function $\mathcal{Z}$ can be
expressed as
\begin{equation}
          \mathcal{Z}=\int dE\; e^{-I(E)}=\int dE\; \nu(E) e^{-\beta E} \label{eq;Z}~,
\end{equation}
where $\nu(E)$ is the density of classical states with energy $E$.
Applying the saddle point approximation, which consists in expanding
the action up to quadratic order around the stationary points
$E_{\text{stat}}$, the integration in Eq.~\eqref{eq;Z} can be
evaluated and the result is finite only if $\delta^2 I
|_{E_\text{stat}}>0$.
On the other hand, in the saddle point approximation the energy and
heat capacity can be approximated as $\langle E\rangle \equiv
-\partial \ln \mathcal{Z}/\partial \beta \approx E_{\text{stat}}$ and
$C\equiv \partial \langle E \rangle / \partial \beta ^{-1}
\approx \partial E_{\text{stat}}/\partial \beta ^{-1}$ and, finally, it
turns out to be
	\begin{equation}
		C\approx \beta^2 \left( \left.\delta^2 I \right|_{E_{\text{stat}}}\right)^{-1}.
\end{equation}
Consequently, since a configuration of a thermodynamic system is locally
stable when its heat capacity is positive, one concludes that the convergence
of the integral in the partition function is equivalent to the thermal
stability of the system~\cite{Brown:1994su}.

From this argument we can infer the stability property of the configurations
depicted in Fig.~\ref{fig:2}. In fact, negative (positive) slopes of the
isocharge curves $\bar{\beta}(x)$ are equivalent to positive (negative) heat
capacity. Thus, for the case $Q>0$ (right panel of Fig.~\ref{fig:2}), when two
black hole solutions are allowed at a given temperature, the smallest is
thermodynamically unstable and the largest is stable.  
Instead, for $Q<0$ (left panel of
Fig.~\ref{fig:2}), when three black holes configurations with the same
$\bar{\beta}$ and $Q$ are possible, the smallest and largest are stable while
the intermediate is unstable. This unstable solution thus corresponds to a
maximum of the Euclidean action whose value is associated to the tunnelling
rate between the two stable configurations~\cite{Gross:1982cv}.

Therefore, the result presented in Fig.~\ref{fig:2} might have striking
implications for cosmology. By taking $r_+\ll r_{\mathcal{B}}$ we expect to
mimic the black hole conditions in the present Universe, since the Universe
is much larger than any event horizon radius. In such a case Fig.~\ref{fig:2}
shows that black holes with $Q\ge 0$ have high temperature and tend to
evaporate. Their relic abundance is then constrained by the usual
astrophysical bounds valid for radiating black holes~\cite{constraints}.  On
the contrary, black holes with $Q<0$ are cold long-lived configurations, which
in general are poorly constrained and may be plausible dark matter
candidates~\cite{constraints}. Of course, the discovery of such stable objects
would be an important finding in favor of massive gravity. Moreover, even
though massive gravity can qualitatively explain either dark energy and dark
matter by tuning the gravitational force at long and intermediate
distances~\cite{darkmatter}, the presence of
black hole energy density might be a fundamental ingredient to fit the data.
\begin{figure*}
\begin{center}
  \hspace{-.2cm}\subfigure{\includegraphics[scale=0.8]{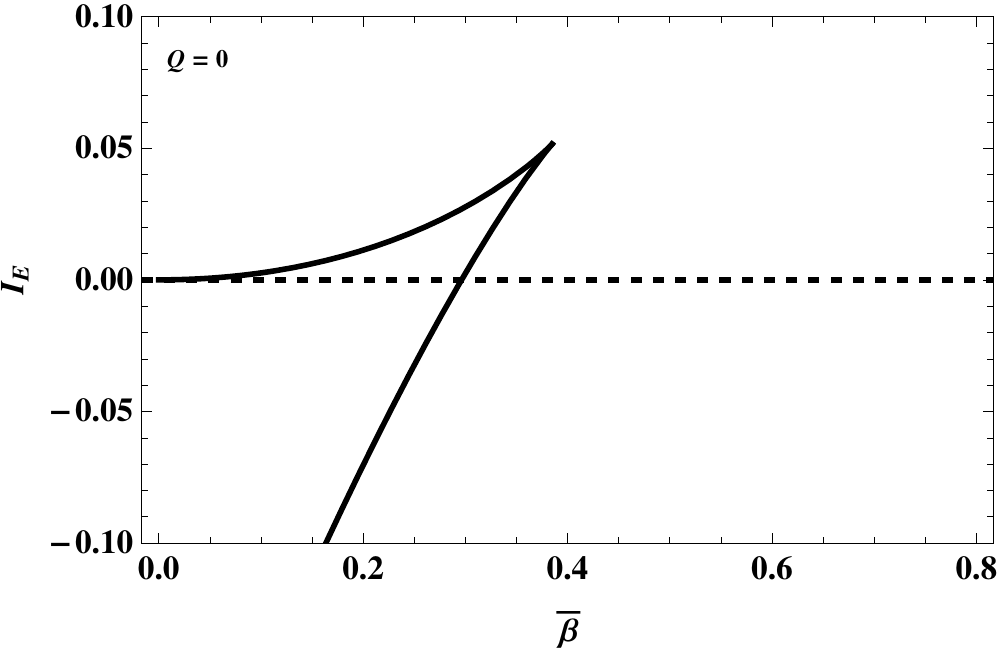}}\hspace{1.cm}\subfigure{\includegraphics[scale=0.8]{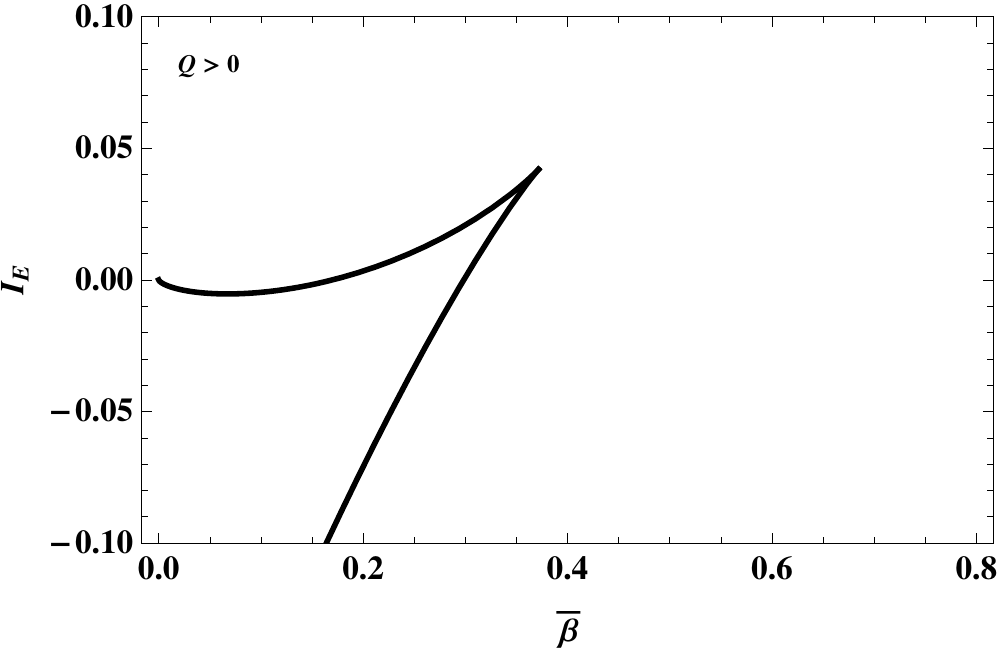}}
   \caption{The Euclidean action as a function of $\bar{\beta}$ in the case
     $Q=0$ (left panel) and $Q>0$ (right panel). The solid and dashed lines
     correspond to a black hole and vacuum phase, respectively.}
   \label{fig:5}
   \end{center}
\end{figure*}
As a last remark, we highlight that for locally-stable black hole
configurations, the action $I_E$ fulfills the three requirements of
Sect.~\ref{sec:1}. Thus, for such configurations $\mathcal{Z}$  can
be interpreted as the partition function of the system. From
Eq.~\eqref{eq:onsh} one can therefore compute thermodynamical quantities as,
for instance, the entropy:
	\begin{equation}
		S=\beta \left(\frac{\partial I_E}{\partial \beta}\right)-I_E=\frac{1}{4}\mathcal{A}_{\text{BH}}~,
\end{equation}
which is the usual Bekenstein-Hawking formula
\cite{key-0.1} and agrees with the result found in
Ref.~\cite{arXiv:1102.0479} by means of Wald's formula
\cite{Wald:1993nt}.  Hence, the Euclidean action reduces to the free
energy in the semiclassical approximation, i.e.~$I_E=\beta F$ with
$F=M-T_H S$, so that the first law of black hole thermodynamics is
recovered by requiring $F$ to have a minimum. Moreover,
as demonstrated in Ref.~\cite{Brown:1989fa} for static
spherically symmetric gravitational systems, the thermal energy is
identical to the quasilocal energy:
	\begin{equation}
		\mathcal{E}_{BY}=\langle E \rangle= \frac{\partial I_E}{\partial \beta}.
\end{equation}

\subsection{Global stability, phase structure and critical
  behavior} 
\label{sec:global}

In the previous section we have determined when the black hole solutions
\eqref{eq:sol} are locally stable. In particular, we found that under certain
conditions multiple locally-stable solutions are allowed. However, for some
fixed values of $Q$ and $\beta$, only one of these solutions can be globally
stable (i.e. it corresponds to the global minimum of $I_E$) while the others
have to be metastable (i.e. they are local but not global minima of $I_E$).
\begin{figure*}
\begin{center}
\hspace{-.6cm}\subfigure{ \includegraphics[scale=0.8]{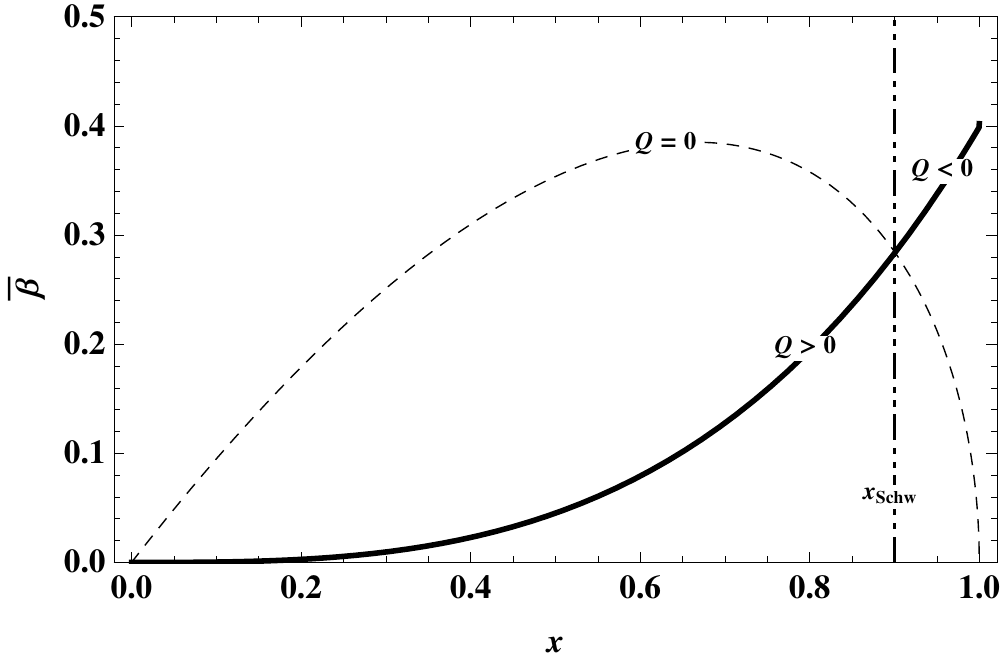}}\hspace{.9cm}
       \subfigure{ \includegraphics[scale=0.8]{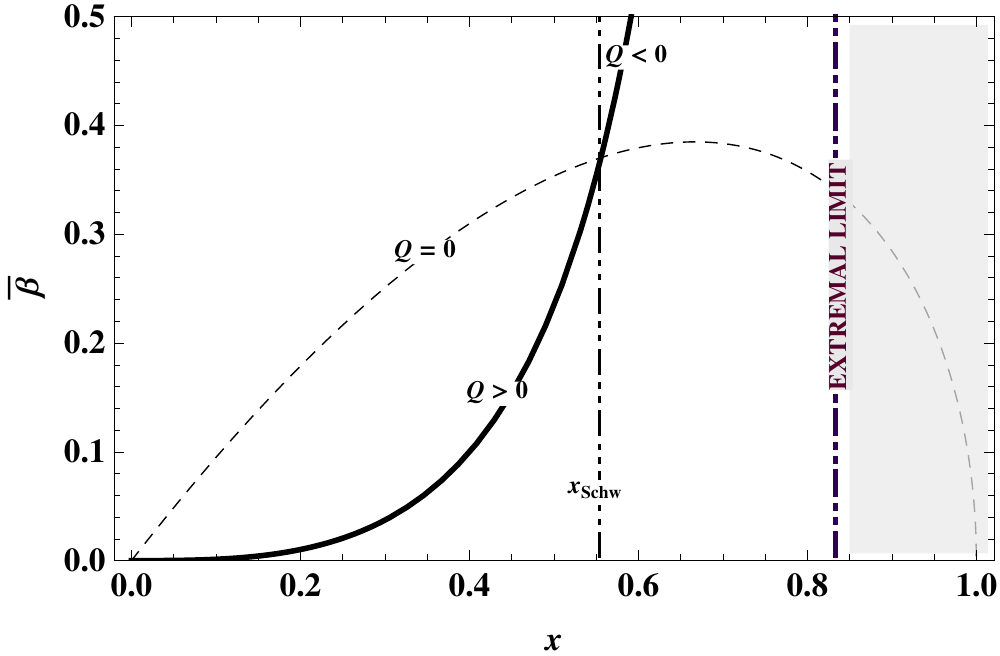}}\\
\hspace{-1.15cm}
\subfigure{ \includegraphics[scale=0.8]{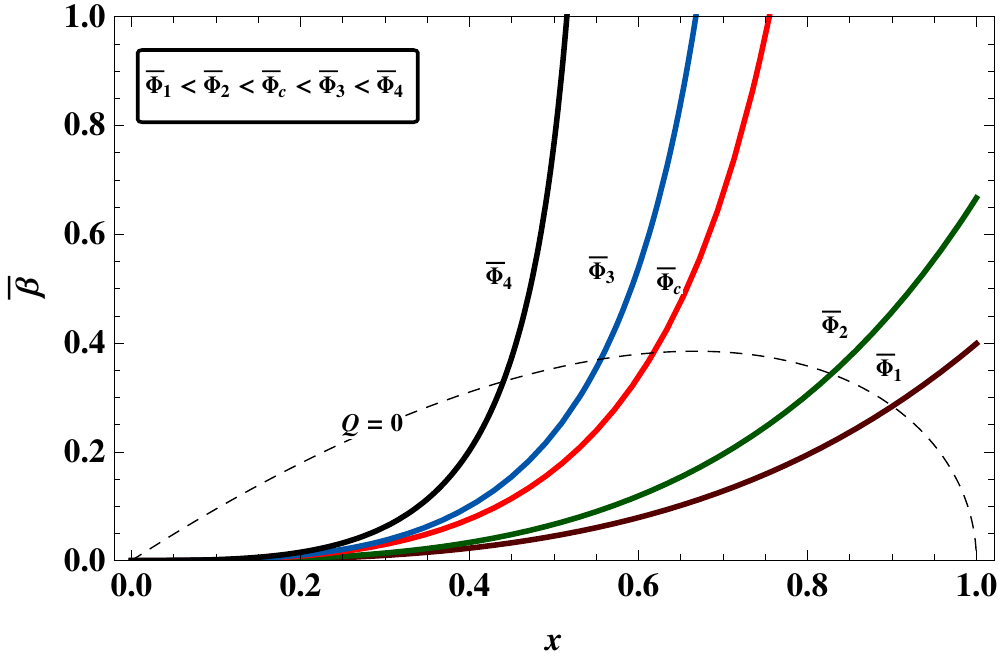}}
     \caption{\underline{Top left panel}: $\bar{\beta}$ as a function of the event horizon for a
       particular value of $\bar{\Phi}<\bar{\Phi}_{\rm c}$. At
       $x=x_{\text{Schw}}$ the scalar charge of the solution changes
       of sign. \underline{Top right panel}: $\bar{\beta}$ as a
       function of the event horizon for a value of
       $\bar{\Phi}>\bar{\Phi}_{\rm c}$. The black hole never engulfs
       the cavity and tends to the extremal limit at low
       temperatures. For $x$ higher than the extremal limit, the
       configuration has no event horizon. \underline{Bottom panel}:
       $\bar{\beta}$ as a function of the event horizon for different
       values of $\bar{\Phi}$.}
   \label{fig:6}
   \end{center}
\end{figure*}

In order to make manifest the phase structure of the black hole solutions
\eqref{eq:sol}, we analyze $I_E$ as a function of temperature for a fixed
$\QQ$.  The result is shown in Fig.~\ref{fig:4} and~\ref{fig:5}, respectively for negative
and positive scalar charge. To read off the informations the
figures contain, it can be useful to take into account also Fig.~\ref{fig:2}.
When $Q$ is negative and $|\mathcal{Q}|\geq|\mathcal{Q}_{\text{crit}}|$ (top right
and bottom panels of Fig.~\ref{fig:4}), there is only one solution
independently of the specific temperature, as we already inferred from
Fig.~\ref{fig:2}.  Instead, when $Q$ is negative but
$|\mathcal{Q}|<|\mathcal{Q}_{\text{crit}}|$ (top left panel of Fig.~\ref{fig:4}),
the competition between the black hole phases is more curious. Starting at very
low temperature and then heating up the system we see that at the beginning only one
solution exists.  This phase remains globally stable up to the critical
temperature $\bar{\beta}_c^{-1}$ ($\bar{\beta}_c\simeq 0.33$ in the example of
the Fig.~\ref{fig:2} and \ref{fig:4}) above which it becomes
metastable. Subsequently, a first order phase transition may occur and,
interestingly, above a certain temperature ($\bar{\beta}\lesssim 0.17$ in the
example) it has to. This property of the phase transition holds
also starting at high temperature and then cooling down the
system\,\footnote{This feature might have interesting applications, as for
  instance in Randall-Sundrum or QCD models, which typically suffer from a too
  long-lived metastable phase~\cite{Rattazzi}.}.
Such a phase structure is isomorphic to the case of Reissner-Nordstrom black
hole in AdS \cite{Chamblin:1999tk} and in a cavity \cite{Lundgren:2006kt} for
the canonical ensemble (fixed electric charge). To test the similarity, we
examine in detail the behavior of our case at
$\mathcal{Q}=\mathcal{Q}_{\text{crit}}$ near the critical
temperature $T_c=\bar{\beta}_c^{-1}$. The entropy and the heat capacity scale as (see Appendix for details)
\begin{eqnarray}
		S-S_{\text{c}}&\propto& \left(T-T_{\text{c}}\right)^{1/3}~\,,\\
		C&\propto&  \left(T-T_{\text{c}}\right)^{-2/3},
\end{eqnarray}
implying that there is a second order phase transition at the critical temperature
$T_c$. The value of  the critical exponent of the heat capacity
has thus the same value as for Reissner-Nordstrom black holes in AdS, dS and
flat space in a cavity at fixed electric charge
\cite{Chamblin:1999tk, Carlip:2003ne, Lundgren:2006kt}.
Such a universal behavior is remarkable, knowing that the Euclidean action is
different.

In some sense, the phase structure for $Q<0$ and
$|\mathcal{Q}|<|\mathcal{Q}_{\text{crit}}|$ looks like the one for
$Q=0$ (left panel of Fig.~\ref{fig:5}) apart from two main
differences: {\it i)} at high temperature the allowed phase is not a
black hole solution but the vacuum spacetime \eqref{eq:2} (dashed line
in the figure); {\it ii)} starting at low temperatures and then
heating up the system, the first order phase transition (possible at
$\bar{\beta}\lesssim0.3$ in the figure) is not always guaranteed since
the vacuum phase is allowed at high temperature\,\footnote{Although
  the presence of the scalar fields, such a phase structure is very
  similar to the one arising in GR for Schwarzschild black holes
  enclosed in a cavity \cite{York:1986it} and Schwarzschild-AdS black
  holes~\cite{Hawking:1982dh}.}. Instead, for $Q>0$ (right panel of
Fig.~\ref{fig:5}) the phase structure is different: no phase is
allowed at low temperature ($\bar{\beta}>\bar{\beta}_{\rm max}\simeq
0.38$ in the example).  A possible explanation of this result might be
that at low temperature the existing phase is a state that has scalar
charge and no event horizon. This could perfectly be a star with low
free energy \cite{Pilo:2008,Bebronne:2009mz}, which could not emerge from our
analysis.  On the other hand, one might guess that this peculiarity
arises because we are assuming that $Q$ is conserved. In the next
section, we assume other boundary conditions where the scalar charge
is allowed to vary inside the cavity.

\section{Other boundary conditions}\label{sec:5}
In principle, one can assume the scalar charge not to be conserved inside the
cavity. In such a case the phase evolution is no longer constrained by the
initial value of $Q$ but by the  value of $\Phi$, the scalar-charge
``potential" at the surface of the cavity. In this ensemble it is straightforward
to see that the Euclidean action compatible with the regularity condition
\eqref{eq:T} is given by
	\begin{equation}
		I_E=\beta \left(\mathcal{E}_{BY}+\Phi Q \right) -\frac{1}{4} \mathcal{A}_{\text{BH}}~, \label{eq:b2}
	\end{equation}
where 
	\begin{equation}
		\Phi =\frac{1}{2 \sqrt{\alpha(r_{\mathcal{B}})}} \left[r_+^{1-\lambda}-r_{\mathcal{B}}^{1-\lambda}\right]~. \label{eq:phi}
\end{equation}
One may prove that the present ensemble is always dominated by a configuration with a negative heat capacity and is therefore ill-defined, as we shortly see now.

By combining Eqs.~\eqref{eq:m1} and \eqref{eq:phi}, $\beta$ can be expressed
as function of $r_+$ and $\Phi$. This relation can be more conveniently
rewritten as
	\begin{widetext}
	\begin{equation}
		\bar{\beta}=\frac{x^{2+\lambda } \left(1-x^{\lambda-1 }\right)^2 \bar{\Phi} }{x^2 (-1+\lambda )-2 x^{1+\lambda } (-1+\lambda )-x^{3 \lambda } \bar{\Phi} ^2+x^{2 \lambda } \left(-1+\lambda +\bar{\Phi} ^2+(-1+x) \lambda  \bar{\Phi} ^2\right)},
\end{equation}
\end{widetext}
where $x\equiv r_+/r_{\mathcal{B}}$, $\bar{\beta}\equiv\beta/4\pi
r_{\mathcal{B}}$ and $\bar{\Phi}\equiv2\Phi
r_{\mathcal{B}}^{\lambda-1}$. The behavior
of the solution is plotted in Fig.~\ref{fig:6} (solid lines) for
several values of $\Phi$. The scalar charge changes along the curve
and it is positive (negative) when the solution is below (above) the
dashed line. The radius at which this cross occurs is marked as
$x_{\text{schw}}$ in Fig.~\ref{fig:6} (top panels, the left plot
being for $\bar{\Phi}<\bar{\Phi}_{\rm c}$ and the right one for
$\bar{\Phi}>\bar{\Phi}_{\rm c}$). Furthermore, when 
$\bar{\Phi}>\bar{\Phi}_{\rm c}$, the black hole event horizon radius
is bounded from above by the extremal case limit and can never reach
the spherical cavity (see Appendix for some analytic results).

The behavior  of $\bar{\beta}$ described above seems problematic
since the positiveness of the heat capacity depends on the sign of the slope
of $\bar{\beta}$. For a positive slope, as in the case here, the heat capacity
is negative. As we have discussed in Sect. \ref{sec:local}, this necessarily
implies a divergent integral for the partition function, i.e.~the partition
function in the saddle point approximation is not well-defined and produces an
imaginary result. This may be interpreted as the effect of an unidentified
metastable phase. However, the present framework does not offer an unambiguous
description and such interpretation should be taken with
precaution.

\section{Conclusions}\label{sec:6}

Lorentz breaking massive gravity is an interesting
theory that may explain the recent acceleration of
the Universe without invoking dark energy.  It also provides peculiar
black hole solutions due to the presence of hair parameters that
modifies the standard gravitational potential. In this
theory the analog of the Schwarzschild black hole --
the asymptotically flat spherically symmetric solution -- depends on
two parameters: the mass and the ``scalar charge'' (characterizing the
hair strength). In this paper we have analyzed equilibrium states and
phase structures of such a solution enclosed in a spherical surface
kept at a fixed temperature.

We have proven that when the scalar charge inside the cavity is not
conserved, the ensemble is ill-defined. On the contrary, when the
scalar charge is held fixed, 
the saddle point approximation can be applied to obtain the
partition function $\mathcal{Z}=e^{-I_E}$, where $I_E$ is the
regularized on-shell Euclidean action. With that formalism we were
able to study the black hole thermodynamics and phase structure. In
particular, the black hole entropy follows the Bekenstein-Hawking formula.

Depending on the value of the scalar charge $Q$ contained in the
cavity, the phase structure presents completely different behaviors
that can be summarized as follows (being $Q_{\text{crit}}$ a critical
value of the scalar charge):

\begin{itemize}

\item{\it $Q> 0$:} Above a certain temperature there exist
  two black hole solutions with different event horizon radii. The
  smaller black hole is unstable while the larger is globally stable
  and describes the phase present in the cavity. Below that
  temperature we do not find any solution with event horizon, probably
  because the analysis is not suitable for this phase.
\item{\it $Q=0$:} Below a certain temperature no black
  hole solution exists and the phase consists of the usual
  globally-flat background of massive gravity. Above that temperature
  the previous phase competes with a new phase consisting of a large
  black hole.
\item{\it $Q_{\rm crit}<Q<0$:} There is an interval of
  temperatures where there are three black holes solutions with
  different sizes of event horizon. The smallest and the largest are
  locally stable and therefore correspond to local minima of the free
  energy, but the intermediate is unstable as it corresponds to a
  maximum of the free energy. Hence, a tunnelling between the two
  stable phases may occur  and this
  first order phase transition is forced to happen when the
  temperature crosses the whole interval.
\item{\it $Q=Q_{\rm crit}$:} Only one black hole solution
  is allowed at each temperature. At a certain temperature a
  second-order phase transition happens and at this moment the
  critical exponent of the heat capacity is $-2/3$.
\item{\it $Q<Q_{\rm crit}$:} There is a single (both
  globally and locally) stable black hole at every temperature.
\end{itemize}

Further investigations on the cases with negative $Q$ are
worthwhile. Indeed, their isomorphism to Reisser-Nordstrom black holes
in AdS, dS and flat space in the canonical ensemble
\cite{Chamblin:1999tk, Carlip:2003ne, Lundgren:2006kt} is peculiar,
knowing that their Euclidean actions are different. Furthermore, at
low temperature the globally stable configurations are small cold
black holes. These objects overcome most of the astrophysical
constraints~\cite{constraints} and, in principle, might be the dark
matter candidates of massive gravity. Dedicated dark matter analyses
would be needed to check this possibility but, as a first step, one
should understand whether the black hole scalar charge is actually
conserved (or at least varies very slow) in nature. In order to
address this question one should probably comprehend the origin of the
modified character of black hole horizons
\cite{Dubovsky:2007zi,key-2}, which is hard to understand without a
known UV completion of the theory. However, with the use of the
AdS/CFT correspondence one may attempt to study such problems
\cite{Niarchos:2009qb}. In that context, the study of the phase
structure may play an important role as a valuable test of the AdS/CFT
correspondence. Since the phase structure emerging for AdS boundary
conditions appears as well as for asymptotically-de Sitter black holes
and asymptotically-flat black holes in cavities~\cite{Carlip:2003ne},
one should expect the CFT side to have a rich variety of phases (such
as deconfinement/confinement) dual to what has been studied in this
paper. Moreover, Lorentz symmetry violation should arise in the CFT
side~\cite{Sundrum:2007xm}. Trying to understand regimes in which
Lorentz violation is sizeable will hopefully allow us to devise the
right experimental tests to decide whether massive gravity is or
is not realized in nature.

\acknowledgments We are grateful to P.~Tinyakov for stimulating
discussions and useful advices. The work of F.C. is
supported in part by the IISN, Belgian
Science Policy (under contract IAP V/27) and by the
``Action de Recherche Concert\'{e}s" (ARC), project ``Beyond Einstein:
fundamental aspects of gravitational interactions" .

\appendix

\section{Some analytic results for $\lambda=2$}

In this appendix, we consider the case $\lambda=2$. This particular
choice allows us to determine the main feature of the general case by
some analytic calculations.  In Appendix~\ref{app:1} we focus on the
ensemble with fixed scalar charge $Q=Q_{\text{crit}}$ and in
Appendix~\ref{app:2} we determine the behavior of $\bar{\beta}$ for
the ensemble with fixed scalar charge potential.

\subsection{Case with fixed scalar charge $Q$}
\label{app:1}

For negative values of the scalar charge the function $\bar{\beta}$
has two extrema (see left panel of Fig.~\ref{fig:2}): one maximum and
one minimum. The location of such extrema is obtained by solving the
equation
	\begin{equation}
	5 \mathcal{Q}^2+(2-2 \mathcal{Q}-3 x) x^3-6 \mathcal{Q} x (-1+\mathcal{Q}+x)=0~. \label{eq:A}
	\end{equation}
        The critical scalar charge is obtained when both extrema are
        degenerate, i.e. when the discriminant of the polynomial
        equation \eqref{eq:A} is vanishing. By performing the explicit
        computation, the discriminant of the equation \eqref{eq:A}
        takes the form:
	 \begin{equation}
 	\triangle=	-\xi_0\QQ^4\sum_{n=3}^{9}\xi_n \QQ^n~,
	\label{eq:discr}
 	\end{equation}
where $\xi_i$ are real and positive constants. 	
 Therefore, several values for $\QQ$ are allowed for a vanishing discriminant \eqref{eq:discr}. However, only one value of $\QQ$ corresponds to a positive event horizon radius: 
 	\begin{equation}
		\mathcal{Q}_{\text{crit}}=\frac{1}{-9-4 \sqrt{5}}~.
	\end{equation}
        By replacing the critical value of the scalar charge in
        Eq.~\eqref{eq:A}, we obtain the critical value of the event
        horizon radius:
 	\begin{equation}
		x_{\text{crit}}=5-2 \sqrt{5}~.
\end{equation} 
By substituting both $x_{\text{crit}}$ and $\mathcal{Q}_{\text{crit}}$
in $\bar{\beta}$, we have the critical inverse temperature
 	\begin{equation}
 		\bar{\beta}_{\text{crit}}= \frac{5}{2} \sqrt{85-38 \sqrt{5}}~.
 	\end{equation}
We note that such critical values are similar to the Reissner-Nordstrom case \cite{Carlip:2003ne}. At $\mathcal{Q}=\mathcal{Q}_{\text{crit}}$, we can expand $\bar{\beta}$ around the critical point $x_{\text{crit}}$: 
	\begin{equation}
		\bar{\beta}-\bar{\beta}_{\text{crit}}=\frac{1}{3!}\left. \frac{\partial ^3 \beta}{\partial x^3}\right|_{x=x_{\text{crit}}} (x-x_{\text{crit}})^3+\dots~. 
	\end{equation}
The  second derivative of $\bar{\beta}$ vanishes, when evaluated at the critical point. From here, one can very easily obtain the behavior of the entropy and heat capacity near the critical temperature: 
	\begin{eqnarray}
		S-S_{\text{crit}}&\propto& \left(T-T_{\text{crit}}\right)^{1/3},\\
		C&\propto&  \left(T-T_{\text{crit}}\right)^{-2/3}.
\end{eqnarray}
As we have already mentioned in Sect.~\ref{sec:global}, the critical
exponent for the heat capacity has the same value $-2/3$ as for
Reissner-Nordstrom black holes in AdS, dS and flat space in a cavity
at fixed electric charge \cite{Chamblin:1999tk, Carlip:2003ne,
  Lundgren:2006kt}. This implies that at the critical value of the
scalar charge, we have a second order phase transition.  Moreover,
this conclusion is independent of the value of $\lambda$ since the
point $x_{\text{crit}}$ is a stationary point of inflection as one can
see from Fig.~\ref{fig:2}.

\subsection{Case with fixed scalar charge potential $\Phi$}
\label{app:2}

Depending on the value of the scalar charge potential, we have two qualitatively different behaviors for $\bar{\beta}$:
\begin{itemize}

\item for $0<\bar{\Phi}< \bar{\Phi}_c=1$: the extremal black hole limit is not allowed inside the cavity and the event horizon radius can therefore take all the possible values inside the cavity, i.e. $0<x<1$. At some temperature, the black hole engulfs the cavity.

\item for $\bar{\Phi}\geq\bar{\Phi}_c=1$: the black hole event horizon is bounded from above by the extremal black hole limit that takes place at $x=1/\bar{\Phi}$. Therefore, at very low temperatures the ensemble is dominated by nearly extremal black holes. In such a case, the black hole can never reach the cavity wall. 

\end{itemize}
In both situations, depending on the value of the event horizon radius, we can have positive or negative scalar charge configurations. The change of sign of $\QQ$ takes place at 
	\begin{equation}
		x_{\text{schw}}=-\frac{1}{2\bar{\Phi}^2}+\frac{1}{2}\sqrt{\frac{1+4\bar{\Phi}^2}{\bar{\Phi}^4}}~,
	\end{equation}
        which is the value where the scalar charge vanishes, turning
        the hairy black hole into the conventional Schwarzschild black
        hole. Positive scalar charge configurations are in the domain
        where $x<x_{\text{schw}}$, implying therefore that black holes
        with negative scalar charge have $x>x_{\text{schw}}$.

At high temperatures, the ensemble is dominated by black holes with positive
scalar charge. As the temperature drops, the scalar charge becomes smaller and
the event horizon becomes larger.

{}

\end{document}